\documentclass[preprint,showpacs,preprintnumbers,amsmath,amssymb]{revtex4}

\usepackage{graphicx}
\usepackage{dcolumn}
\usepackage{bm}
\usepackage[german,american]{babel}

\begin{document}

\title{Properties of ideal Gaussian glass-forming systems}

\renewcommand{\thefootnote}{\fnsymbol{footnote}}

\author{Andreas Heuer}
\author{Aimorn Saksaengwijit*}

\affiliation{University of M\"unster, Institute of Physical
Chemistry, Corrensstr. 30, D-48149 M\"unster, Germany \\ *King
Mongkut's University of Technology Thonburi, Thailand}

\date{\today}
\begin{abstract}

We introduce the ideal Gaussian glass-forming system as a model to
describe the thermodynamics and dynamics of supercooled liquids on
a local scale in terms of the properties of the potential energy
landscape (PEL). The first ingredient is the Gaussian distribution
of inherent structures, the second a specific relation between
energy and mobility. This model is compatible with general
considerations as well as with several computer simulations on
atomic computer glass-formers. Important observables such as
diffusion constants, structural relaxation times and kinetic as
well as thermodynamic fragilities can be calculated analytically.
In this way it becomes possible to identify a relevant PEL
parameter determining the kinetic fragility. Several experimental
observations can be reproduced. The remaining discrepancies to the
experiment can be qualitatively traced back to the difference
between small and large systems.

\end{abstract}


\maketitle

\section{Introduction}

The understanding of the dynamics of supercooled liquids is still
far from being complete
\cite{Debenedetti01,Binder:2005,Dyre:2006,Lubchenko:2006}. A lot of
insight has been gained from simulations . For example, in real
space the microscopic nature of dynamic heterogeneities has been
clarified
\cite{Hurley:1995,Kob:92,Donati:98,Donati:94,Donati:355,Heuer:89,Qian:249}.
Using the framework of the potential energy landscape (PEL) a lot of
insight could be also gained in configuration space
\cite{Wales:2003,Sciortino05}. A key aspect is the use of inherent
structures (IS) , i.e. local minima of the PEL. Upon minimization
basically all configurations can be mapped on a IS. In this way the
regular dynamics can be mapped on a hopping dynamics between IS
\cite{Stillinger:222,Stillinger:245}. Physically, this mapping can
be interpreted as a removal of the vibrational degrees of freedom.
However, as explicitly shown in \cite{Schroder:210} the properties
of the structural relaxation remain identical for sufficiently low
temperatures. Generally speaking, the mapping on the IS can be
interpreted as a coarse-graining procedure. At low temperatures the
IS dynamics displays many correlated forward-backward jumps between
adjacent IS. In a further coarse-graining step it is possible to
define metabasins (MB) by an appropriate merging of adjacent IS
\cite{Stillinger:1995,Middleton:214,H9,H10,Denny:2003}. In this way
the effect of correlated forward-backward motion has basically
disappeared.

A key question deals with the relation between thermodynamics and
dynamics. For example the empirical Adam-Gibbs relation
\cite{Adam65}
\begin{equation}
\label{ag} \Gamma(T) = \Gamma_0 \exp(-B_{AG}/Ts_{c}(T))
\end{equation}
relates the configurational entropy $s_c$ to the local relaxation
rate $\Gamma$. A further relation between thermodynamics and
dynamics is formulated via the fragilities.  In the spirit of the
thermodynamic fragility as discussed in \cite{Martinez2001,Wang06}
one can define the thermodynamic fragility index via
\cite{Ruocco:2004}
\begin{equation}
m_{thermo} = -\beta_g \frac{S_c^\prime(\beta_g)}{S_c(\beta_g)}.
\end{equation} where $T_g = 1/\beta_g$ (choosing $k_B = 1$) denotes the
glass-transition temperature.  Furthermore, the kinetic fragility
is defined as
\begin{equation}
\label{defmkin}
 m_{kin} = d \ln \tau_\alpha /d(T_g/T).
\end{equation}
Qualitatively, it denotes the slope of the relaxation time (or
viscosity) in the Angell-plot \cite{angell,Martinez2001}.
Empirically, one finds a significant correlation between the
kinetic and the thermodynamic fragility \cite{Martinez2001}. In
principle the kinetic fragility may also be defined for the
diffusion constant. Due to the violation of the Stokes-Einstein
relation \cite{Fujara92} minor variations of the value of
$m_{kin}$ will be present. Furthermore, it turns out that for the
set of all glass-forming systems one observes a significant
correlation between $m_{kin}$ and the degree of
non-exponentiality, expressed, e.g., by the exponent $\beta_{KWW}$
of the stretched exponential function \cite{Bohmer:279}. If one
restricts oneself, however, to the set of all molecular
glass-forming systems (excluding in particular network forming
systems and polymers) the residual correlation is very weak
(-0.28) and the values of $\beta_{KWW}$ are restricted for most of
the systems ($> 80\%$) in that work to a relatively small regime
between 0.5 and 0.62 \cite{Bohmer:279,review}. In contrast, the
network-forming systems are characterized by nearly exponential
relaxation and small values of $m_{kin}$.

In the language of the IS or the MB the thermodynamic properties
at constant volume are to large extent determined by their energy
distribution $G(e)$.  For many systems it has been shown
numerically that the distribution of IS can be described by a
Gaussian \cite{Sciortino:1999,H6,Starr:140,Nave04}. Even for
BKS-SiO$_2$ the distribution is Gaussian, albeit displaying a
low-energy cutoff in the range of accessible temperatures for
computer simulations \cite{H17}. Furthermore, it turns out that
the distribution of IS and MB is nearly identical in the relevant
regime of low-energy states \cite{H10}.

Within the PEL approach it is possible to relate the thermodynamic
and the dynamic aspects \cite{H9,H10,Denny:2003,H13}. This is
based on the observation that the escape rate from a MB can be
expressed in terms of its energy, i.e. $\Gamma(e,T)$. Furthermore,
it turns out that the temperature dependence of the diffusion
constant $D(T)$  can be fully expressed in terms of the average
local escape rate. As a consequence, knowledge of $G(e)$ and
$\Gamma(e,T)$ allows one to predict $D(T)$.  A similar type of
relation between energy and mobility can be found, e.g., for the
trap model \cite{Monthus:310}.

The goal of this work is to elucidate the properties of a system
with a Gaussian distribution $G(e)$ of MB. The functional form of
$\Gamma(e,T)$  is rationalized by different models, discussed in
literature, and at the same time by comparison with previous
computer simulations on the binary mixture Lennard-Jones system
(BMLJ) and silica (BKS-SiO$_2$). On this basis we define an {\it
ideal Gaussian glass-former} (IGGF). For the IGGF several
observables can be determined analytically such as the temperature
dependent diffusion constant and relaxation time, its kinetic and
thermodynamic fragility and its non-exponentiality. In this way it
becomes possible, e.g., to elucidate the relevant PEL parameters
which determine the fragility. In Sect.II the IGGS is introduced
and in Sect.III its main properties are calculated. We end with a
critical discussion and a summary in Sect.IV.

\section{Description of the ideal Gaussian glass-former}

\subsection{Thermodynamics}

Of crucial importance for the properties of a glass-forming system
is the number density of IS, denoted $G(e)$. Here we always
consider a system with $N$ particles. For many different systems,
studied via computer simulations, a Gaussian density of IS has
been found \cite{Starr:140, Nave04,Sciortino:280,H7}, i.e.
\begin{equation}
\label{Gauss_g} G(e)  =\exp(\alpha N) \frac{1}{\sqrt{2\pi
\sigma^2}} \exp(-(e-e_0)^2/2\sigma^2).
\end{equation}
A notable exception is BKS-SiO$_2$. This system is characterized
by a low-energy cutoff \cite{H22} which gives rise to the
fragile-to-strong crossover \cite{Saika:2004,H22}. In principle,
for the calculations, shown below, the effect of a low-energy
cutoff can be incorporated \cite{review}. Here we mainly
concentrate just on the case of a purely Gaussian density of IS.

\begin{figure}
\includegraphics[width=6cm]{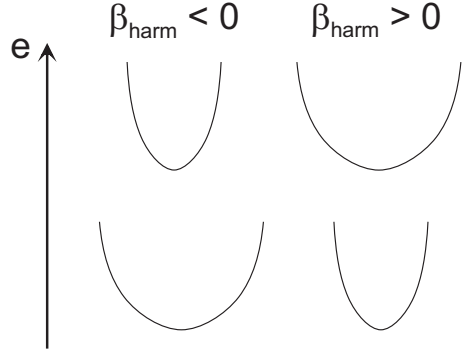}
\caption{\label{betaharm} A sketch of the cases $\beta_{harm} < 0$
and $\beta_{harm} > 0$. Shown are typical curvatures around
representative IS at different energies.}
\end{figure}

For a closer discussion one has to take into account that the
average curvature around the minima may depend on $e$. For different
systems it turns out to a very good approximation that one has a
linear energy-dependence for the free energy $F_{harm}(e)$, related
to the harmonic vibration in a well
\cite{Sciortino:1999,H6,Sastry:198,Starr:140,Mossa:284,Gio2003,Sciortino2003}.
This can be written as
\begin{equation}
\label{betaharme}
 F_{harm}(e) = const - \beta_{harm}e.
\end{equation}
The constant $\beta_{harm}$ is a material constant. The meaning of
the sign of $\beta_{harm}$ is visualized in Fig.\ref{betaharm}.

The Boltzmann distribution $p_{eq}(e)$ describes the probability to
be (at a randomly chosen time) in an IS with energy $e$.
 $p_{eq}(e)$ is proportional to
$G(e) \exp(-\beta e)$ when $\beta_{harm} = 0$. Taking into account
the curvature-effect, introduced above, one finds
\begin{equation}
\label{bharm} p_{eq}(e) \propto G_{eff}(e) \exp(-\beta e)
\end{equation}
 with the
effective density
\begin{equation}
G_{eff}(e) \propto G(e) \exp(-F_{harm}(e)) \propto
\exp(-(e-e_{0,eff})^2/2\sigma^2).
\end{equation}
and $e_{0,eff} = e_0 + \beta_{harm} \sigma^2$. Thus the presence
of an energy-dependent average curvature can be incorporated by a
shift of the Gaussian distribution of states.

The standard definition of the configurational entropy is $-\sum_i
p_i \ln p_i$ where the sum is over all states (not energies).
Mapping this relation to the description in terms of energies one
obtains
\begin{equation}
\label{entropy}
 S_c(T) = \int de p_{eq}(e) \ln (G(e)/p_{eq}(e)) = \int de
p_{eq}(e) S_c(e) - \int de p_{eq}(e) \ln p_{eq}(e).
\end{equation}
For the $G(e)$ and $p_{eq}(e)$, obtained for the Gaussian
distribution, one obtains from the first term
\begin{equation}
\label{entropys} S_c(T) = N \alpha - (1/2) \sigma^2 \left (\beta -
\beta_{harm} \right )^2.
\end{equation}
 For large $N$ one expects that $\sigma^2 \propto N$ due to the
central limit theorem. Then $S_c(T)$ becomes extensive as
expected. In contrast, the last term in Eq.\ref{entropy}, which
would give rise to $1/2$, can be neglected because it is not
extensive and just gives rise to a minor redefinition of $\alpha$
($\alpha \rightarrow \alpha + 1/2N$).

Defining the Kauzmann-temperature by the condition $S_c(T_K) = 0$
(and $\beta_K = 1/T_K$) Eq.\ref{entropys} can be equivalently
expressed as
\begin{equation}
T S_c(T) = [(N \alpha) + \sigma^2 \beta \beta_K/2 - \sigma^2
\beta_{harm}^2/2](T - T_K).
\end{equation}
Neglecting the temperature-dependence of the second term this is
actually the standard expression when deriving the VFT-temperature
dependence (e.g. $\ln (D(T)/D_0) \propto 1/(T - T_0)$ where often
$T_0 \approx T_K$ is found) from the Adam-Gibbs expression
\cite{Sastry:198}. Using a similar way of rewriting the
configurational entropy, this type of argument can be already
found in \cite{Sastry:198}. In any event, for the further analysis
we will use the expression Eq.\ref{entropys} due to its
simplicity.

\subsection{Transitions between MB: Models}

There is a long history of models which describe the dynamics in
configuration space on a phenomenological level
\cite{Brawer:1984,Dyre:1987,Arkhipov:1994,Monthus:310,Diezemann:1997,Diezemann:1998}.
One considers a region of the viscous fluid  which can
cooperatively rearrange via a transition state. For the time being
the initial and final states may be characterized by the energy of
the respective IS (or MB). For sufficiently low temperatures the
elementary rearrangement process is considered to be activated:
the system leaves a state with energy $e$, crosses a high-energy
transition state with rate $\Gamma(e)$ (from now on the variable
$T$ is omitted) and ends up in a new state which is uncorrelated
to the initial one. Different names can be found for essentially
identical models (e.g. trap model, free energy model) following
this scenario.

\begin{figure}
\includegraphics[width=6cm]{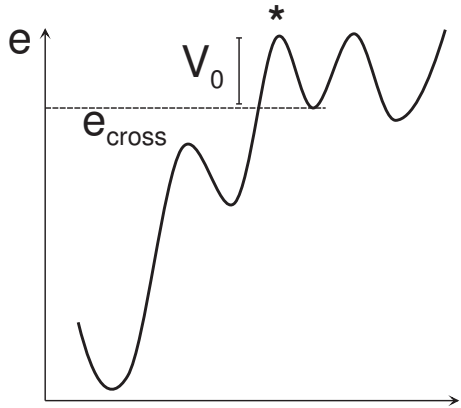}
\caption{\label{ecrossdef} Sketch of the multi-step escape
process, including the definition of $V_0$. The barrier with the
star is supposed to be the critical barrier beyond which $p_{back}
< 0.5$.}
\end{figure}

The hopping rate $\Gamma(e)$ is characterized by two energies.
$e_{cross}$ denotes the energy of the IS just after the final
barrier, which has a height $V_0$; see Fig.\ref{ecrossdef} for the
sketch. According to the model assumptions $e_{cross}$ and $V_0$
are independent of the initial energy $e$.  Actually, even in more
complex systems like the random energy model one can argue via
percolation arguments that $e_{cross}$ is independent of $e$
\cite{Dyre95}. More generally, in a percolation-like picture of
the PEL $e_{cross}$ corresponds to the energy level from which on
the system finds adjacent states with similar energies and thus
does not have to increase further in the PEL for the final
relaxation. Defining $E_{app}(e)$ as the apparent activation
energy to escape from energy $e$, this scenario can be written as
\begin{equation}
\label{gammae} \Gamma(e) = \Gamma_0(e) \exp(-\beta E_{app}(e))
\end{equation}
with
\begin{equation}
E_{app}(e) = e_{cross}+V_0 - e
\end{equation}
for $e \le e_{cross}$ and $E_{app}(e) = V_0$ for $e \ge
e_{cross}$. Stated differently, the escape for energies lower than
$e_{cross}$  is solid-like (activated) whereas otherwise it is
liquid-like \cite{H13}.

The energy-dependent prefactor $\Gamma_0(e)$ reflects possible
entropic effects. As argued in \cite{Brawer:1984,Brawer:1985} the
prefactor $\Gamma_0(e)$ contains an energy-dependent factor
$M_{entro}$ which denotes the number of escape paths to reach a
high-energy state with energy $e_{cross}$. In most models this is
neglected by simply choosing $\Gamma_0(e) = \Gamma_0$. This would
be justified in case of 1D reaction paths or low-dimensional
percolation paths. A simple expression for $M_{entro}$ can be
formulated if every state with energy $e_{cross}$ can be reached
from exactly one state with energy $e (< e_{cross})$. It is given
by $M_{entro} = G(e_{cross}) / G(e)$ \cite{H20}, i.e.
\begin{equation}
\label{g0start} \Gamma_0(e) = \Gamma_0 G(e_{cross}) / G(e).
\end{equation}
This holds for $e < e_{cross}$ in the opposite limit one just has
$\Gamma_0(e) = \Gamma_0$.  For $e_{cross} - e \ll e_0 - e_{cross}$
Eq.\ref{g0start} can be approximated as
\begin{equation}
\label{gamma0orig} \Gamma_0(e) \approx \Gamma_0
\exp((e_0-e_{cross})(e_{cross} - e)/\sigma^2).
\end{equation}
For later purposes this is rewritten as
\begin{equation}
\label{gamma0_theo} \Gamma_0(e) = \Gamma_0 \exp(\kappa
k_{entro}(e_{cross} - e))
\end{equation}
with
\begin{equation}
\label{kentrodef}
 k_{entro} =  (e_{0,eff}-e_{cross})/\sigma^2
 \end{equation}
and
\begin{equation}
\label{kappadef}
 \kappa =  \frac{e_{0}-e_{cross}}{e_{0,eff}-e_{cross}}
 \end{equation}

\begin{figure}
\includegraphics[width=6cm]{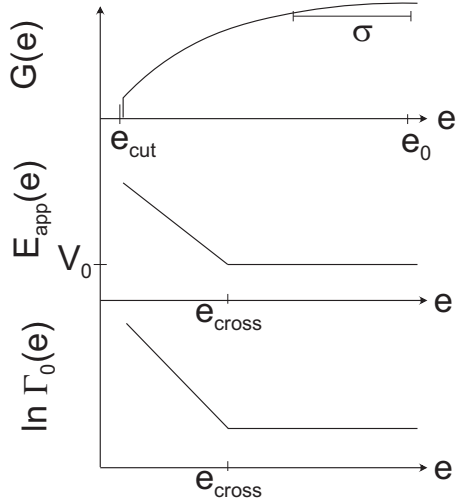}
 \caption{\label{ealldef} Sketch of the energies, introduced in the text. A possible difference
     between $G(e)$ and $G_{eff}(e)$ is neglected.}
\end{figure}

This somewhat complicated way to rewrite Eq.\ref{gamma0orig} is
motivated in two ways. First, because $p_{eq}(e)$ is directly
related to $G_{eff}(e)$ and thus to $e_{0,eff}$, see
Eq.\ref{bharm}, it is more convenient to use $e_{0,eff}$ rather
than $e_0$. Second, in practice the factor $\kappa$ has to be
treated as an empirical parameter because the increase of the
entropic term $\Gamma_0(e)$ with decreasing energy may somewhat
deviate from the specific scenario, described above. The relevant
energy scales are summarized in Fig.\ref{ealldef}.

It is convenient to introduce the shifted inverse temperature
\begin{equation}
\label{betastardef} \tilde{\beta} = \beta - \kappa k_{entro}.
\end{equation}
In principle all calculations, shown in this work, can be
performed as well for $\kappa \ne 1$. However, since the influence
of the entropic prefactor is not as important as the energetic
term the additional complexity of the expressions is not worth the
additional information for $\kappa \ne 1$. In what follows we
therefore always choose $\kappa = 1$.

Now one can rewrite Eq.\ref{gammae} as
\begin{equation}
\label{centralgamma} \Gamma(e) = \Gamma_0
\exp(-\tilde{\beta}(e_{cross}-e)) \exp(-\beta V_0)
\end{equation}
and, for a Gaussian density of states,
\begin{equation}
\label{peq_Gauss} p_{eq}(e) \propto  G_{eff}(e) \exp(-\beta e)
\propto \exp[-(e-e_{cross}+\sigma^2\tilde{\beta})^2/2\sigma^2].
\end{equation}

When comparing Eq. \ref{centralgamma}  with simulations one has to
take into account that the simulated system may contain more than
one elementary system. Each subsystem is characterized by an energy
$e_i$ and $e = \sum e_i$. For a superposition of  $M$ {\it
independent} subsystems the total hopping rate $\Gamma_M(e)$ is just
the sum of the individual hopping rates $\Gamma(e_i)$. To a first
approximation one may assume that the energy $e$ is equally
distributed among the $M$ subsystems, yielding $\Gamma_M(e) = M
\Gamma(e/M)$. A closer analysis shows that apart from another
energy-independent factor this is indeed the correct expression
\cite{review}. This expression for $\Gamma_M(e)$ suggests to
generalize Eq.\ref{centralgamma} to
\begin{equation}
\label{centralgamma2} \Gamma(e) = \Gamma_0 \exp(-\lambda
\tilde{\beta}(e_{cross}-e)) \exp(-\beta V_0)
\end{equation}
Here $1/\lambda$ is a measure for the number of elementary
subsystems, present in the specific system. This completes the
definition of the IGGF.  For later purposes we introduce the
dimensionless quantity
\begin{equation}
\mu \equiv \tilde{\beta} \sigma
\end{equation}
which will turn out to be the central quantity characterizing the
properties of the IGGF.

\subsection{Comparison with simulations}

The above scenario has been compared with simulations of
relatively small systems of the BMLJ system $(N=65)$ \cite{H10}
and BKS-SiO$_2$ $(N=99)$ \cite{H20}). This comparison has been
performed for MB in order to have a random-walk type dynamics in
configuration space. With IS it would have been impossible to
express observables such as the diffusion constant or the
relaxation time just in terms of the waiting times
\cite{H10,review}. For the comparison the average waiting time in
MB of a given energy $e$ have been determined,  denoted as
$\langle \tau(e) \rangle$. Naturally, the average escape rate
$\Gamma(e)$ is then given by
\begin{equation}
\Gamma(e)= \frac{1}{\langle \tau(e) \rangle}.
\end{equation}
Note that this definition does {\it not} imply that the escape
from a MB with energy $e$ corresponds to an exponential waiting
time distribution with average waiting time $\langle \tau(e)
\rangle$.

The  simulations have fully confirmed the validity of
Eq.\ref{centralgamma2} except for a slight smearing out effect for
energies close to $e_{cross}$. Actually, the effective barriers
could be identified by a closer analysis of the relevant minima
and saddles of the PEL \cite{H10}. Actually, in \cite{Souza:2006}
it has been shown that the additional barrier before the final
transition (denoted $V_0$ above) and the barrier, governing the
local forward-backward motion at low temperatures (within a MB)
are roughly the same.

Very recently, de Souza and Wales have analyzed the temperature
dependence of the mean square displacement, evaluated for a fixed
time $\tau$ \cite{Souza:2006}. Of course, for very large $\tau$
this analysis recovers the standard diffusion coefficient. For
ambient $\tau$, which for the lowest temperatures is significantly
shorter than $\tau_\alpha$, the authors observe a simple Arrhenius
behavior with the high-temperature activation energy $V_0$. For
lower temperatures this approach is sensitive to the local
forward-backward motion within a MB. The barriers in this regime
are of the order of $V_0$ so that the local processes remain
activated with the high-temperature activation energy. This
strengthens the observation that it is roughly the same value
$V_0$ which governs the additional barrier height at low and high
energies.

Furthermore it turns out that $\Gamma_0(e)$ indeed shows an
exponential dependence of energy. Interestingly,
$\Gamma_0(e_{cross}) \approx 1/20$ fs$^{-1}$ is of the order of
typical molecular time scales. This also suggests that the
increase below $e_{cross}$ is due to entropic reasons.

The PEL parameters, obtained from the fitting, are listed in
Tab.\ref{tab1}. Note that if not mentioned otherwise from now on
all energies are expressed relative to $e_{0,eff}$, i.e. the
maximum of $G_{eff}(e)$. For the analytical calculations, to be
presented below, it is convenient to exclusively use
Eq.\ref{centralgamma2}, i.e. using $e < e_{cross}$ and
$\tilde{\beta} > 0$. The first relation starts to be very well
fulfilled if $e_{cross} - \langle e(T) \rangle
> \sigma$ which roughly implies $T < 0.6$ in case of
BMLJ and $T < 3600$ K in case of BKS-SiO$_2$. In this temperature
range one also has $\tilde{\beta} > 0$.

\begin{table}
\centering
  \begin{tabular}[t]{|c|c||c|c|c|c||c|c|c|c|c|}\hline
 & & \multicolumn{4}{|c||}{thermodynamic} &  \multicolumn{5}{|c|}{dynamic} \\ \hline
   & N &  $\sigma $ & $  - e_{cut}$ & $\beta_{harm}$ & $\alpha $ &  $  - e_{cross}$ & $ \lambda $ &
$\kappa$ &  $V_0$ & $\Gamma_0$
       \\ \hline
      BKS-SiO$_2$ & 99 & 3.5 eV& 43.4 eV & $\approx 0$ & 1.14 & 37.5 eV & 0.66 &   0.62  &  0.8 eV & 1/(20 fs) \\ \hline
      BMLJ  & 65 &  3.0  & - & -0.3 & 0.73 & 12.9 &  0.55 & 0.3 &  1.0 & 1/150  \\ \hline
       \end{tabular}
       \caption{ The thermodynamic and dynamic PEL parameters,
obtained from simulation of BKS-SiO$_2$ and BMLJ.} \label{tab1}
\end{table}

Interestingly, $e_{cross}$ is significantly smaller than
$e_{0,eff}$. As will
 become clear below this difference is crucial for properties like
 the fragility.
The additional barrier height $V_0$ is present both for
BKS-SiO$_2$ and BMLJ (and has similar height after normalization
by $\sigma$). Therefore $V_0$ cannot be of any relevance for the
question of fragility. It can be directly extracted from the
high-temperature behavior.

The observation $\lambda < 1 $ suggests than even these small
systems are not elementary.  This is equivalent to the result
reported in \cite{Denny:2003} that a consistent mapping on an
elementary trap model is not possible.

Two major differences are evident when comparing BKS-SiO$_2$ and
BMLJ. First,  the low-energy cutoff for BKS-SiO$_2$ is
significantly larger than the cutoff, dictated by entropy. Thus,
the amorphous ground-state is a finite-entropy state. Second,
 $(-
e_{cross})/\sqrt{\lambda \sigma^2}$ is much lower for BKS-SiO$_2$.
This means that activated processes become relevant only for
states much lower in the PEL. As a consequence, a characteristic
temperature like $T_{MCT}$ should be lower for silica than for
BMLJ. Indeed, $\Delta (\sigma/ T_{MCT}) \equiv
(\sigma/T_{MCT})_{silica}-(\sigma/ T_{MCT})_{BMLJ}\approx 12.2-6.7
= 5.5$ \cite{LaNave:2002,Doliwa_Eacc:2003} and $ \Delta
(-e_{cross})/\sigma) = 6.4$ are similar. Furthermore, the
energy-dependence of $\Gamma_0(e)$ for BKS-SiO$_2$ is much more
prominent.

\section{The dynamics of ideal Gaussian glass-forming systems}

\subsection{General}

The MB dynamics can be characterized by a waiting time
distribution $\varphi(\tau)$ \cite{H10}. From this one can
calculate the different moments  $\langle \tau^n \rangle$ of
$\varphi(\tau)$.  It has been shown in previous work that the
diffusion constant $D$ is proportional to $1/\langle \tau \rangle$
\cite{H9}. Within the continuous-time random walk (CTRW) formalism
the structural relaxation time $\tau_\alpha$ can be identified
with $\langle \tau^2 \rangle/\langle \tau \rangle$
\cite{Berthier:2005a}. Actually, very recently it has been shown
\cite{Rubner_tbp,review} that it is indeed fully justified to use
the CTRW-formalism to describe the dynamics of the  BMLJ ($N=65$)
system.

Given the distribution of energies as well as the relation between
energy and mobility one may ask whether one can explicitly calculate
$\langle \tau^n \rangle$. For this purpose we first introduce
$\varphi(e)$ as the probability density that in a series of
different MB, visited by the system, a randomly chosen MB has energy
$e$. Then the average waiting time is given by averaging $\langle
\tau(e) \rangle$ over all MB, i.e.
\begin{equation}
\label{tau_phi} \langle \tau \rangle = \int \, de \, \varphi(e)
\langle \tau(e) \rangle = \int \, de \, \varphi(e) / \Gamma(e).
\end{equation}

$\varphi(e)$ is distinctly different from the Boltzmann
distribution $p_{eq}(e)$ which denotes that at a randomly given
time the present MB has energy $e$, i.e. $p_{eq}(e) \propto
\varphi(e) \langle \tau(e) \rangle$. Including a normalization
factor this can be rewritten as
\begin{equation}
\label{phi_vs_p}
 p_{eq}(e) = \frac{\varphi(e)}{\Gamma(e) \langle
\tau \rangle}.
\end{equation}
Qualitatively, this relation expresses that low-energy states
(small $\Gamma(e)$) are often observed (at randomly chosen times)
although their actual number $\propto \varphi(e)$ may be very
small. Multiplication of Eq.\ref{phi_vs_p} with $\Gamma(e)$ and
subsequent integration yields
\begin{equation}
\label{tau_vs_p}
 \langle \tau \rangle^{-1} = \int de \, p_{eq}(e) \Gamma(e) \equiv \langle \Gamma \rangle_p.
\end{equation}
Thus, the average waiting time is also related to the rate average
over the equilibrium probability distribution. Note the different
notations ($\langle . \rangle$ as the $\varphi$-average vs. $\langle
. \rangle_p$ as the $p$-average.  Using the explicit form of
$G_{eff}(e)$ one obtains after a straightforward integration
\begin{equation}
\label{tau-1} \langle \tau \rangle^{-1}= \Gamma_0
\exp((\lambda^2/2 - \lambda) \mu^2/2) \exp(-\beta V_0)
\end{equation}

So far no information about the nature of the relaxation process
has entered the analysis. In the most simple case the escape from
a state with energy $e$ is governed by a single barrier height.
Then the waiting time distribution, related to this energy, is
just $\Gamma(e) \exp(-\Gamma(e) t)$. For the BMLJ(N=65) system one
has $1/\lambda \approx 2$ subsystems. In the most simple picture
the total energy is then the sum of two independent subsystems,
each with energy $e_i$ ($e_1 + e_2 = e$) and for a given energy
decomposition the total rate $\Gamma(e)$ is given by $\Gamma(e_1)
+ \Gamma(e_2)$. Actually, as outlined in \cite{review}, the
normalized second moment $\langle \tau(e)^2 \rangle / \langle
\tau(e) \rangle^2$ is expected to be around 16 for $T=0.5$ for 2
subsystems as compared to 2 for an elementary system. The
broadening of the waiting time distribution at fixed energy is due
to the fact that for a given total energy $e$ several
decompositions $e = e_1 + e_2$ are possible, each giving rise to
different escape rates. The numerically observed value is approx.
8 \cite{H18}. This means that the BMLJ(N=65) system behaves, to
first approximation, like two independent subsystems (each
described by $\lambda = 1$ and variance $\sigma^2/2$ if $\sigma^2$
is the variance of the original system). A possible reason for the
decrease of 16 to 8 will be given below. In any event, in what
follows we neglect this effect and postulate that the elementary
system behaves like an IGGF with $\lambda = 1$ and an exponential
waiting time distribution at given energy. Since the waiting time
distribution at fixed energy is a well-defined observable in the
MB approach the subsequent calculations could be easily
generalized to take into account deviations from a purely
exponential behavior of the waiting time distribution of the
elementary system.

This aspect is strongly related with the old discussion of
homogeneous vs. heterogeneous relaxation
\cite{Schmidt-Rohr,Richert93}. Heterogeneous relaxation would simply
mean that one has a superposition of exponentially relaxing
entities. Experimentally it has been shown that the dynamics at the
glass transition is basically heterogeneous \cite{Bohmer:football}.
This indicates that the choice of an exponential waiting time
distribution is indeed not too bad.

\subsection{Calculation of moments}

With this approximation the waiting time distribution
$\varphi(\tau)$ and the distribution $\varphi(e)$, reflecting the
thermodynamics, are related via
\begin{equation}
\label{waiting_t_d} \varphi(\tau) \propto \int de \, \int dt \,
\varphi(e) \exp(-\Gamma(e) t) \delta(t - \tau).
\end{equation}
Its different moments $\langle \tau^n \rangle$ can be directly
calculated
\begin{equation}
\label{time_n} \langle \tau^n \rangle  =  \int de \varphi(e) n!
\Gamma(e)^{-n}  =  n!\langle \tau \rangle \langle
\Gamma^{1-n}\rangle_p \exp(n \beta V_0).
\end{equation}
For the second equality Eq.\ref{phi_vs_p} has been employed.

Straightforward evaluation of Gaussian integrals yields
\begin{equation}
\label{tau_star} \langle (\Gamma/\Gamma_0)^m \rangle_p = \exp [(
m^2/2 -  m)\mu^2 - m \beta V_0].
\end{equation}

The case $m=1$ recovers Eq.\ref{tau-1}. Furthermore, the case
$m=-1$ gives finally rise to
\begin{equation}
\label{tau-2} \langle \tau^2 \rangle / \langle \tau \rangle^2 =
\exp(\mu^2)
\end{equation}

In most models no distinction between $e_{cross}$ and $e_{0,eff}$
is made. Then $\mu$ can be identified with $\beta$. The relations
for this special case can be already found in literature
\cite{Monthus:310}. Note that in this limit Eq.\ref{tau-1}
corresponds to the well-known $1/T^2$ temperature dependence,
discussed, e.g., in \cite{Dyre95}.

\section{Applications}

\subsection{Kinetic fragility}

\begin{figure}
\includegraphics[width=6cm]{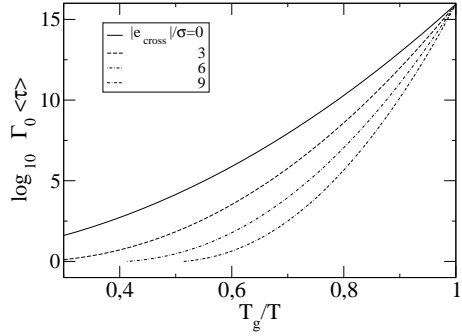}
\caption{\label{fragile1} The temperature dependence of $ \Gamma_0
\langle \tau \rangle (\propto D(T))$ for different values of the
crossover energy with $\lambda = 1$ (the values are given with
respect to $e_{0,eff}$).  }
\end{figure}

Here we analyse the temperature dependence of $\langle \tau
\rangle$ (and thus of $D(T)$) and in particular the fragility. The
glass transition temperature is defined by the condition
\begin{equation}
\Gamma_0 \langle \tau(T_{g,K}) \rangle  = 10^{K}.
\end{equation}
Neglecting for the time being the somewhat different temperature
behavior of $D(T)$ and $\eta(T)$ (see below) $T_{g,16}=
1/\beta_{g,16}$ roughly corresponds to the calorimetric $T_g$
because $\eta(T_g) / \eta(T \gg T_g) \approx 10^{16}$. Simple
expressions emerge for the case $V_0 = 0$ (corrections can be
simply calculated but only mildly influence the results). Using
Eq.\ref{tau_star}  a simple calculation yields
\begin{equation}
\label{beta_g} \sigma \beta_{g,K} =  k_{entro} \sigma
+\sqrt{2K\ln(10)}.
\end{equation}
In relation to the definition of $T_g$ we use the notion
$m_{kin,K}$ rather than $m_{kin}$ (see Eq.\ref{defmkin}) to
express the dependence on the time scale. Then a straightforward
calculation yields
\begin{equation}
\label{fragility}
 m_{kin,K} =  2K  + \sqrt{2K/\ln(10)}k_{entro} \sigma .
\end{equation}
In this regime the fragility depends on the dimensionless
parameter $k_{entro}\sigma = - e_{cross}/\sigma$. Thus, the
dynamic crossover energy is a central PEL parameter determining
the fragility. These results are visualized in Fig.\ref{fragile1}.
One can clearly see how the fragility increases with increasing
$-e_{cross}/\sigma$.

Note that Eq.\ref{fragility} implies that BMLJ is stronger than
BKS-SiO$_2$ if the cutoff were artificially removed so that the
PEL is purely Gaussian. The non-fragile behavior of BMLJ has been
already mentioned in Ref.\cite{Tarjus:2000}.

Of course, since the temperature dependence of $\tau_\alpha$ is in
general not identical to that of $\langle \tau \rangle$ the
results would slightly differ if $m_{kin,K}$ is calculated for
$\tau_\alpha$ or $\eta$ rather than for the diffusivity.

Empirical relations to correlate the fragility with, e.g., the
Poisson ratio have been suggested \cite{Novikov:2004} but are
questioned in \cite{Johari06}. It would be interesting to check
whether there exists a physical connection between the
observables, suggested in that work, and the value of $e_{cross}$,
determining the crossover from liquid-like behavior to solid-like
behavior.

\subsection{Relation to the AG approach}

Alternatively, one can calculate the value of $\beta_g$ under the
assumption of the AG relation Eq.\ref{ag} and a Gaussian PEL
(using $\beta_{harm} = 0$). Then one has to solve the equation
\begin{equation}
10^K  = \exp(\beta_g B_{AG}/(\alpha - \beta_g^2 \sigma^2/2N).
\end{equation}
For large $K$ one obtains
\begin{equation}
\label{beta_ag} \beta_g = \frac{\sqrt{2\alpha N}}{\sigma} -
\frac{B_{AG}\sqrt{N}}{\sigma K \ln (10)}.
\end{equation}
Then a straightforward calculation yields for the fragility (again
in the limit of large $K$)
\begin{equation}
\label{frag_ag} m_{kin,K} = \frac{\sqrt{2\alpha} K^2 (\ln 10)^2
\sigma}{B_{AG}\sqrt{N}}.
\end{equation}
Within the AG-approach the fragility depends on the density of
states, i.e. $\alpha$, as well as the empirical constant $B_{AG}$.
A large number of states implies larger fragility (at least for
fixed $B_{AG}$ which, of course, could also depend on $\alpha$
\cite{Ruocco:2004}).

It may be interesting to compare this relation with the fragility
Eq.\ref{fragility}, obtained for an IGGF. Qualitatively, both
relations would show a somewhat similar behavior if systems with
large $\alpha$ are related to systems with a low crossover energy
$e_{cross}$, i.e. large $k_{entro}$. This is not unreasonable
because in the spirit of percolation-like arguments for a larger
number of IS the system would be able to find a path with a lower
barrier to move between two low-energy IS. However, in a strict way
it will not be possible to map Eq.\ref{frag_ag} on
Eq.\ref{fragility} because of the different $K$-dependence.
Formally, this problem could be solved if $\alpha$ decreases with
increasing $K$, i.e. going to longer time scales and thus lower
glass transition temperatures. Qualitatively this statement is
equivalent to the requirement that $G(e)$  decays faster than a
Gaussian. This has been suggested in \cite{Matyushov2005}.
Physically this might, e.g., occur as a consequence of a broadened
low-energy cutoff.

\subsection{Thermodynamic fragility}

In the spirit of  the thermodynamic fragility as discussed in
\cite{Martinez2001,Wang06} one can define the thermodynamic
fragility index via \cite{Ruocco:2004}
\begin{equation}
m_{thermo,K} = -\beta_g \frac{S_c^\prime(\beta_g)}{S_c(\beta_g)}.
\end{equation}
We obtain, using Eq.\ref{entropys},
\begin{equation}
m_{thermo,K} = \frac{\sigma^2 (\beta_g -
\beta_{harm})\beta_g}{N\alpha - \sigma^2(\beta_g -
\beta_{harm})^2/2}.
\end{equation}
Note that the denominator must be positive because otherwise the
entropy of the system would be negative. Under this condition, an
increase of $\sigma k_{entro}$ (which is the only relevant
dimensionless parameter, characterizing IGGF)  and thus of $\sigma
\beta_g$ (via Eq. \ref{beta_g}), gives rise to an increase of
$m_{thermo}$ and $m_{kin}$, independent of the values of $\beta_g$
or $\alpha$. This strong correlation of $m_{kin}$ and $m_{thermo}$
is in agreement with the experimental observation for most systems
\cite{Martinez2001}.

Interestingly, increasing the value of $\beta_{harm}$ yields a
decrease in $m_{thermo,K}$. However, a different behavior emerges
if one includes the vibronic contribution into the entropy, i.e.
by using $S_{ex}(T) = S_{c}(T) + S_{harm}(T)$ rather $S_c(T)$. A
straightforward calculations yields $S_{ex}(T,\beta_{harm}) =
S_c(T,-\beta_{harm})$, thereby neglecting a constant and a term,
depending logarithmically on $\beta$ \cite{review}. Accordingly,
when defining $m_{kin,K}$ on the basis of $S_{ex}(T)$ one obtains
an increasing thermodynamic fragility for increasing
$\beta_{harm}$ in agreement with the qualitative discussion in
\cite{Martinez2001}.

If the cutoff starts to influence the system a detailed
calculation is no longer possible because the behavior of the
configurational entropy at low temperatures depends on the details
of $G(e)$ at low energies. Thus, it is not surprising that for
SiO$_2$ the thermodynamic fragility does not follow the general
trend \cite{Martinez2001}.

The present discussion complements the work by \cite{Sastry:198}
where the kinetic and the thermodynamic fragility have been
discussed with reference to the AG-relation. Simulations have also
revealed a significant correlation between both fragilities.

\subsection{Relaxation properties}

Here we ask for the probability $S_0(t)$ that a system in
equilibrium has not performed a hopping process until time $t$. It
is given by
\begin{equation}
\label{S01} S_0(t) =  \int de \, p_{eq}(e) \exp(-\Gamma(e) t).
\end{equation}
In what follows the trivial factor $\exp(\beta V_0)$ will be
omitted. For sufficiently low temperatures the decay of this
function can be related to the structural relaxation
\cite{Berthier:2005a, Rubner_tbp}.

As shown in \cite{Castaing:1991,review} one can approximate for
intermediate times ($S_0(t) \approx 1/e)$
\begin{equation}
S_0(t) \approx \exp(-(t/\tau_{KWW})^{\beta_{KWW}})
\end{equation}
with
\begin{equation}
\label{kww} \beta_{KWW} = 1/\sqrt{1+\mu^2}
\end{equation}
 and $\tau_{KWW} =
1/\Gamma^\star$ where
\begin{equation}
\Gamma^\star = \Gamma(\langle e(T) \rangle ) = \exp(-\mu^2).
\end{equation}

  This may justify the use of the stretched
exponential as a fitting function at least for intermediate times.
This result is insensitive to the specific form of $\Gamma(e)$
since $\Gamma(e)$ only enters via $\Gamma^\star$. Note that for
the IGGF the non-exponentiality tends to increase when going to
lower temperatures. Furthermore one can show that in very
long-time decay is algebraic \cite{Castaing:1991,review}
\begin{equation}
S_0(t)  \propto t^{-u/2\mu^2}.
\end{equation}

One can define the $\alpha$-relaxation time $\tau_\alpha$ via
\begin{equation}
\label{taualpha} \tau_\alpha = \int \, dt \, S_0(t)
\end{equation}
which corresponds to the typical time until a particle jumps for
the first time \cite{Berthier:2005a, Rubner_tbp}. From Eq.
\ref{S01} one immediately obtains (also using Eq.\ref{tau_star})
\begin{equation}
\tau_\alpha = \langle \Gamma^{-1} \rangle_p = (1/\Gamma_0)
\exp(3\mu^2/2).
\end{equation}
This has to be compared with the average hopping time $\langle
\tau \rangle$ (Eq.\ref{tau-1}). One obtains
\begin{equation}
\label{se} \tau_\alpha / \langle \tau \rangle = \exp(\mu^2) .
\end{equation}
Since the left side is proportional to $D \tau_\alpha$ Eq.\ref{se}
expresses the invalidation of the Stokes-Einstein relation for
IGGF. Using the definition of the exponent $a$ via $D(T)
\tau_\alpha(T) \propto \tau_\alpha^a$ , i.e. $\langle \Gamma
\rangle \langle \Gamma^{-1} \rangle \propto \langle \Gamma^{-1}
\rangle^\alpha$ one obtains $a=2/3$. Experimental values are
smaller (e.g. 0.25 for orthoterphenyl \cite{Fujara92} and 0.23 for
TNB \cite{Swallen03}). Thus, the decoupling seems to be too
strong. Qualitatively the strong increase of $\tau_\alpha$ with
decreasing temperature is due to the very long-time tail of
$S_0(t)$.

\section{Discussion and Summary}

The IGGF has been introduced, based on the numerical results for
BMLJ and BKS-SiO$_2$ (except for the low-energy cutoff for
BKS-SiO$_2$) at small system sizes. The general concepts are also
compatible with several models proposed to rationalize the dynamics
of supercooled liquids. Thus, one naturally finds how properties
such as the non-exponentiality are generated.

More specifically, the key conclusions are as follows: 1.) If the
cutoff-energy does not interfere the temperature-dependence of the
dynamics is fully captured by the value of $\mu$ (except for a
trivial $\exp(-\beta V_0)$-term). This means in particular that at
$T_g$ an IGGF has a fixed value of $\mu$, independent of $\sigma
k_{entro}$ and thus independent of its fragility. This implies via
Eq.\ref{kww} that the stretching parameter $\beta_{KWW}$ does not
depend on the fragility if determined exactly at $T_g$. This may
rationalize the weak correlation between $\beta_{KWW}$ and
$m_{kin}$ for the molecular glass-forming systems, as mentioned
above.  Of course, residual fluctuations are expected when the
smaller-order effects of $\lambda$, $\kappa$ and $V_0$ are taken
into account. 2.) The fragility of a system is to a large extent
dominated by the crossover energy $e_{cross}$ relative to the
width of the energy distribution, i.e. $\sigma$. Systems are more
fragile if the crossover from solid-like activated dynamics to
liquid-like non-activated dynamics occurs at low energies,
relative to the width of $G(e)$.  Of course, as soon as the
low-energy cutoff of the PEL comes into play (such as for
BKS-SiO$_2$) the system automatically behaves Arrhenius-like and
thus is classified as strong. This also shows that the fragility
is only partly able to classify a glass-forming system because
already the present discussion shows that there at least two very
different parameters, $e_{cut}$ and $e_{cross}$, which strongly
influence the fragility. 3.)Although the BMLJ data can be fit to
the AG-relation, from a conceptual point of view the IGGF is not
compatible with the AG-relaxation. This can be seen from the
different dependence of the fragility on $K$. On a qualitative
level this discrepancy could be reduced if the distribution of
states decays stronger than a Gaussian at the low-energy end. 4.)
The thermodynamic fragility indeed is correlated with the kinetic
fragility, albeit in a non-trivial way. Again, the systems with a
cutoff-behavior (most notably BKS-SiO$_2$) have to be discussed
separately. 5.) Finally, the IGGF displays non-exponential
relaxation with a non-exponentiality which increases with
decreasing temperature and, in agreement with the experiment,
shows a violation of the Stokes-Einstein relation.

 Conceptually,
the presence of individual relaxation processes naturally is
attributed to small systems, reflecting the typical length scales of
cooperative dynamics during single MB transitions. Thus, any strict
comparison with simulations in the framework of the PEL approach is
conveniently performed with small systems. As shown in previous work
the diffusion constant as well as the thermodynamic properties of
the BMLJ ($N=65)$-system only have very minor finite-size effects
when comparing with the results obtained for much larger systems
\cite{H6,H11,Stariolo06}. However, the structural relaxation time as
the well as the non-exponentiality has somewhat larger finite-size
effects \cite{Stariolo06}. This effect can be understood if one
assumes a specific type of coupling between adjacent subsystems of a
larger system. When some subsystem relaxes it may change the
mobility of the adjacent subsystems \cite{review}. A similar idea
can be already found in \cite{Bouchaud:1996,Monthus:310} and has
been also implemented in the context of the rate memory to explain
the results of multidimensional NMR experiments
\cite{Heuer1995,Sil96,Heuer1997,Diezemann:1997}. In this way the
very immobile regions typically become mobile at some stage and can
relax subsequently. In some sense this idea is also related to the
philosophy of the facilitated spin models where the local mobility
is also influenced by the state of the neighbor spins
\cite{Fredrickson84,Garrahan02,Garrahan:2003}. The coupling between
adjacent subsystems can be formulated such that the diffusion
constant and the thermodynamics does not change whereas the
structural relaxation time, all moments $\langle \tau^n \rangle$ for
$n \ge 2$ and the degree of non-exponentiality decrease upon this
coupling \cite{review}. This might also explain why the second
moment for the BMLJ system is by a factor of 2 smaller than expected
(see above). In particular the exponent $a$, characterizing the
violation of the Stokes-Einstein equation approaches experimentally
relevant values \cite{review}.  However, one of the key results,
namely the utmost relevance of a single dimensionless parameter
$\mu$ would still be valid. In any event, the path back from small
systems to macroscopic systems is one of the challenges for future
work. Using the IGGS as the elementary system for such models is
definitely a reasonable starting point.

We gratefully acknowledge important input from C. Rehwald and O.
Rubner as well as very helpful correspondence with L. Berthier
about this topic.


\end{document}